# DOS-Limited Contact Resistance in Graphene FETs

Kosuke Nagashio* and Akira Toriumi
Department of Materials Engineering, The University of Tokyo,
Bunkyo, Tokyo 113-8656, Japan
E-mail address: nagashio@ material.t.u-tokyo.ac.jp

**Abstract**

Graphene has attracted much attention as one of promising candidates of future high-speed transistor materials because of its high carrier mobility of more than 10,000 cm$^2$ V$^{-1}$ s$^{-1}$. Up to this point, we have focused on the contact properties as performance killers, as a very small density of states in graphene might suppress the current injection from metal to graphene. This paper systematically reviews the metal/graphene contact properties and discusses the present status and future requirements of the specific contact resistivity.

## 1. Introduction

Graphene-based devices are promising candidates for future high-speed field-effect transistors (FETs); a high carrier mobility of ~10,000 cm$^2$ V$^{-1}$ s$^{-1}$ on SiO$_2$ has been reported.[1] A key characteristic of graphene is its linear electronic dispersion relation. This results in (i) a zero effective mass of the carrier with a high Fermi velocity ($v_F$) of ~10$^8$ cm/s and (ii) a very small density of states (DOS) around the Fermi energy ($E_F$).[2,3] In general, the high $E_F$ due to the large carrier density (~10$^{22}$ /cm$^3$) results in a high $v_F$ of ~10$^8$ cm/s in metals. In contrast, the $v_F$ for graphene is not determined by the carrier density but by the slope of the linear electronic dispersion relation. Therefore, although the carrier mobility in the graphene channel itself is high enough, the very small DOS for graphene might suppress the current injection from the metal to the graphene, which might limit the total performance of graphene FETs. In this paper, we review the current understanding of metal/graphene contacts.

## 2. Graphene/Metal Contact

We first consider the ideal case without surface states (interface traps). Figure 1 shows the energy band diagrams for (a) metal/semiconductor, (b) metal/metal, and (c) metal/graphene contacts. When a metal and a semiconductor are brought into contact and are connected by an external wire to form a simple circuit, charges flow between the semiconductor and the metal through the external wire in order for the electrochemical potentials, i.e., the Fermi levels ($E_F$), to line up on both sides at the thermal equilibrium. This builds up the Schottky barrier ($\phi_B = \phi_1 - \chi$),[4] where $\phi_1$ and $\chi$ are the work function for metal 1 and the electron affinity for the semiconductor, respectively. A depletion layer with a length of $W_{dp}$ is formed because of the much smaller carrier density in the semiconductor. In contrast, the metal/metal contact has no potential barrier. Although the carrier is transferred directly through the metal/metal interface to cancel the difference in work functions, the small redistribution of the electron cloud can screen this potential difference because of the large carrier density. In general, the screening length, expressed as $\lambda = \left[4\pi N(E_F)\right]^{-\frac{1}{2}}$, where $N(E_F)$ is the DOS at the $E_F$, is very short in metals (typically a fraction of a nm).[5] Therefore, the vacuum level changes sharply at the metal/metal interface.

The case of interest in this work is the metal/graphene contact, which is very similar to the metal/metal contact, because graphene is treated as a metal due to its lack of a band gap. Figure 2 shows the relation between the DOS and the energy, visualizing the charge transfer process taking place "just" at the metal/graphene interface. The amount of charge transfer gradually decreases from the metal/graphene interface, as shown in Fig. 1(c). A very small amount of electron transfer shifts the $E_F$ significantly. It is known that 0.01 electrons per carbon atom would lower the $E_F$ by 0.47 eV.[6] This charge transfer forms the dipole layer at the interface; the potential difference of the dipole layer is expressed as $\Delta V$. Moreover, the very small DOS around the $E_F$ for graphene increases the screening length. The resulting long charge transfer region is a unique characteristic of the metal/graphene contact. It is noted that $\Delta V$ at the interface depends on the strength of the metal/graphene interaction and is different from $\Delta \phi_{1G} = \phi_1 - \phi_G$, where $\phi_G$ is the work function of graphene. In the case of the metal/metal contact, it is meaningless to separate $D\phi_{12}$ into $\Delta V$ at the interface and the potential drop in the charge transfer region because of the short screening length.

The key feature that originates from the long charge transfer region is the *p-n* junction that appears near the metal/graphene contact. Figure 3 illustrates (a) the schematic of the graphene FET and (b) the band diagram that includes the charge transfer region (hole-doping case) in which the traces of the Dirac point are shown by a broken line, a dash-dotted line and a dash-double-dotted line for different gate voltages ($V_G$).

Moreover, the local resistivity for different values of $V_G$ is also schematically shown as a function of position in Fig. 3(c), where the local resistivity at the p-n junction is considered to be the resistivity ($\rho_{DP}$) at the Dirac point. When a positive $V_G$ is applied to the device, a p-n-p junction is formed near the contact (Fig. 3(b)), resulting in an additional resistance, as shown in Fig. 3(c). In contrast, the increase in the series resistance is negligible for a negative $V_G$ because no p-n junction is formed (p-p-p junction). Consequently, as shown in Fig. 3(d), the asymmetric carrier density dependence on $V_G$ is observed due to the increase in the resistance near *p-n* junctions.

This origin for the asymmetry was first studied by comparing devices with invasive electrodes crossing the whole graphene channel width and external electrodes connected to the side of the graphene channel.[7] It was shown that the electric transport in the channel with the external electrodes is not affected by p-n junctions, unlike the invasive electrodes. The increase in the asymmetry with decreasing channel length also supported the existence of the p-n junction.[8] Moreover, the formation of the charge transfer region leads to band bending, that is, a built-in electric field. The shift in the Dirac point by the deposition of small amounts of metals in the graphene channel is also considered to be due to the effect of the charge transfer.[9] Direct evidence of the built-in electric field was provided by a photocurrent experiment in which photoexcitation near the metal/graphene contact generates electron-hole pairs that can be separated by the built-in electric field to produce a net photocurrent.[10] It is noted that the charge transfer takes place even at the floating condition, i.e., no external wire, which was also confirmed by the generation of the photocurrent near the floating contact.[10] The length of the charge transfer region was reported as ~0.5 μm.[11] However, it has also been pointed out that the charge transfer length might depend on the contact area.[12] A detailed study should be conducted. Finally, it is worthwhile to mention that the modulation doping of graphene from the contacts could possibly control the polarity of the channel when the charge transfer length is longer than the channel length. This is a big advantage because the contact doping does not cause additional Coulomb scattering.

## 3. Contact Resistivity

In the previous section, it was assumed that the charge transfer region ascribed to the small DOS in graphene results in the additional resistance caused by p-n junctions. In this section, we show that the small DOS in graphene results in the high contact resistivity ($\rho_C$) at the metal/graphene contact, even though there is no potential barrier in the band diagram, as shown in Fig. 1(c).

### 3.1 Current flow path and contact resistivity at the metal/graphene contact

Several $\rho_C$ values have been reported, especially for a Ti/Au electrode.[13-15] However, they were described in units of either Ω μm or Ω μm$^2$ because the current flow path at the graphene/metal contact was not revealed. In this review, we reveal for the first time whether $\rho_C$ is characterized by the channel width ($W$) or the contact area ($A=Wd$) by using a multiprobe device with different contact areas. All of the graphene FETs were fabricated on 90 nm SiO$_2$/p$^+$-Si substrates by mechanical exfoliation from Kish graphite. The detailed fabrication method is described elsewhere.[16]

Figure 4(a) shows the four-layer graphene device with six sets of four-probe configurations (#1~#6). Ni was employed as the contact metal. The devices with different contact areas for the source and the drain were fabricated, and the contact area for the voltage probes was kept constant to avoid uncertain effects from the voltage probes. The contact resistance ($R_C$) was calculated by $R_C=1/2(R_{total} - R_{ch} \times L/l)$, where $R_{total}$ is the total resistance between the source and the drain, $R_{ch}$ is the channel resistance between the two voltage probes, $L$ is the length between the source and the drain, and $l$ is the length between the two voltage probes, as shown in Fig. 4(b). Figure 4(c) shows the relationship between the contact area and two types of contact resistivities, $\rho_C^A=R_CA$ and $\rho_C^W=R_CW$, which were extracted by the four-probe measurements. $\rho_C^A$ increases with increasing contact area, whereas $\rho_C^W$ is nearly constant for all of the devices. This indicates that $\rho_C$ is not characterized by $A$ but $W$ instead, i.e., the current flows mainly through the edge of the graphene/metal contact. In other words, the current crowding takes place at the edge of the contact metal.[17]

The current crowding should depend on the contact metal. Figure 5 shows the relationship between the contact resistivities ($\rho_C^W$) and $\mu_{4P}/\mu_{2P}$ for the contact metals Cr/Au, Ti/Au and Ni. The two-probe mobility ($\mu_{2P}$) includes the contact resistance. On the other hand, the four-probe mobility ($\mu_{4P}$) eliminates the influence of the contact resistance. The $\rho_C^W$ for Cr/Au and Ti/Au is typically high and varies over three orders of magnitude, from ~10$^3$ to 10$^6$ Ω μm, whereas a small and uniform $\rho_C^W$ is achieved in Ni. The contact resistivity seems to be independent of the layer number for three kinds of contact metals. These results suggest that the selection of the contact metal is crucially important because the outstanding performance of the graphene channel with $\mu_{4P} >$ 10,000 cm$^2$V$^{-1}$s$^{-1}$ is inevitably obscured by a high $\rho_C$.

Next, to understand the edge conduction in the graphene/metal contact, the current flow path is discussed based on the transmission line model (TLM), as shown in Fig. 6.[15] In an equivalent circuit of the TLM, there are three types of resistance: the sheet resistance of the metal ($R_M^S$), the sheet resistance of graphene ($R_{ch}^S$) and the specific contact resistivity ($\rho_{C\square}$). It should be noted that the unit of $\rho_{C\square}$ is defined as $\Omega\ cm^2$. The edge or area conduction can be presumed by considering the relative magnitudes of $R_M^S$ and $R_{ch}^S$. Because $R_M^S$ is smaller than $R_{ch}^S$, it is assumed that the current flows preferentially in the metal to follow the path of least resistance and that it enters graphene at the edge of the contact. Although a low value of $R_{ch}^S$ might be expected from the high mobility of graphene, this is not the case because of the much smaller carrier density compared with that of the metal.

In reality, however, the current does not flow just at the contact edge line. Thus, it is quite useful to estimate the effective contact distance known as the transfer length ($d_T$). The $d_T$ is approximately characterized by the relative magnitude of $R_{ch}^S$ and $\rho_{C\square}$ as

$$d_T = \sqrt{\frac{\rho_{C\square}}{R_{ch}^S}}, \quad (1)$$

where the metal sheet resistance is neglected.[18] Hereafter, the graphene/metal contact is more accurately described by using both $\rho_{C\square}$ and $d_T$ instead of the edge-normalized $\rho_C$. To quantitatively determine $\rho_{C\square}$, the cross-bridge Kelvin (CBK) structure was used.[19] The rectangular shape of monolayer graphene was prepared by using $O_2$ plasma etching. Three electrodes were made on monolayer graphene. A constant current was imposed between two electrodes on the upper side, while the voltage was measured between two electrodes on the right side, as shown in Fig. 7(a). It is a kind of quasi four-probe measurement. Figure 7(b) shows a schematic drawing of the electric potential along the dotted line in Fig. 7(a). The electrical potential is highest near the contact edge and drops nearly exponentially with the distance, where the $1/e$ distance is indeed defined as $d_T$. The voltage measured from the side is the linear average of the potential over the contact length, $d$. Therefore, $\rho_{C\square}$ can be directly measured by the following simple equation:[19]

$$R_C = \frac{V}{I} = \frac{\rho_{C\square}}{dW}. \quad (2)$$

Figure 7(c) shows $\rho_{C\square}$ as a function of the gate voltage ($V_g$). At a high gate voltage (n=$\sim 5\times 10^{12}\ cm^{-2}$), $\rho_{C\square}$ is $\sim 5\times 10^{-6}\ \Omega\ cm^2$. Furthermore, under the assumption that $R_M^S$ is much smaller than $R_{ch}^S$, the sheet resistance of graphene is required to estimate $d_T$ in eq. (1). Because the sheet resistance of graphene was not available from the two-probe geometry shown in Fig. 7(a), both the low and high mobilities that were measured previously[20] were used for the present analysis. Figure 7(c) shows $d_T$ as a function of $V_G$ by considering the high and low mobility cases. The apparent contact length is ~4 μm, but only ~1 μm is effective for the current transfer in the present experiment. Because $d_T$ exhibits similar values for devices with the same $\rho_{C\square}$, current crowding at the contact edge is always observed for the devices with a contact length longer than $d_T$, as shown in Fig. 4(c). If the contact length becomes shorter than $d_T$, a transition from edge conduction to area conduction will occur.

In the calculation of $d_T$, the $R_{ch}^S$ of the channel region was adopted for the $R_{ch}^S$ underneath the metal contact. As previously reported, the $R_{ch}^S$ underneath the metal contact is strongly dependent on the deposition processes.[12] In fact, the sputtering process of Ti produced considerable defects, which was confirmed by the defect-related D band "through" the sputtered-Ti film (8 nm) observed in micro-Raman spectroscopy. This led to the extremely high $\rho_C^W$, as shown in Fig. 5. In contrast, the D band was not obvious for the thermal evaporation using resistive heating. Therefore, it is likely that the $R_{ch}^S$ of the channel region is different from the $R_{ch}^S$ underneath the metal contact, which should be taken into consideration for more accurate estimation of $d_T$.

3.2 Gate voltage dependence of contact resistivity

Recently, it was reported that the $\rho_C$ obtained using the transfer length method for the Ti/Au and Ni electrodes is independent of $V_G$.[15,21] In this method, all of the graphene/metal contacts are assumed to be equivalent. Although we also used this method, a negative $\rho_C$ value was often extracted when $V_G$ was swept. This is because the above-mentioned method should induce a large estimation error for the case in which there is a big difference between $R_C$ and $R_{sh}$. Moreover, the $V_G$ dependence of $\rho_C$ extracted by the four-probe measurement often results in a negative $\rho_C$ value at one side of the Dirac point, as shown in Fig. 8. This negative $\rho_C$ is not intrinsic, which can be explained by the schematic drawings shown in Fig. 9. When $R_{total}$ and $R_{ch}$ are measured by the four-probe measurement, the Dirac points for both cases are not often consistent, suggesting that the averaged potential minima of the whole channel area ($LW$) for $R_{total}$ and the local channel area ($lW$) for $R_{ch}$ are different [see Fig. 4(b)]. This difference seems to be due to the charge transfer from the metal contact and/or the spatially distributed interaction with the $SiO_2$ substrate. Therefore, the simple extraction of $\rho_C = 1/2W(R_{total} - R_{ch}L/l)$ leads to the large swing of $\rho_C$ with $V_G$, as shown in Fig. 9(a). However, when two Dirac points are almost equal to each other, a mountain-like shape is obtained. This consideration suggests that the $V_G$ dependence of $\rho_C$ cannot be extracted correctly by the conventional four-probe measurement. Therefore, it should be emphasized that the mountain-like $V_G$ dependence of $\rho_{C\square}$ determined using the CBK structure [Fig. 3(c)] is more reliable because the single contact of the graphene/metal interface was directly measured, in contrast with the transfer length method and the four-probe measurement.

This mountain-like $V_G$ dependence of $\rho_{C\square}$ might be explained by the $V_G$ dependence of the DOS for graphene "under the metal contact". Because the DOS in graphene increases with the carrier density, the dependence of $\rho_{C\square}$ on the carrier density may possibly result in a mountain-like shape. However, it is not clear whether the carrier in graphene in contact with the metal with a large DOS is modulated because the electrostatic potential induced by the backgate bias is expected to be screened by the metal on graphene. Although the possibility of carrier modulation in graphene under the metal contact is reported,[22] direct evidence is not available at present. Elucidating the physical properties of graphene under the metal contact is the key to further understanding the metal/graphene contact. It is noted that we have considered no carrier modulation case for graphene under the metal contact, that is, the pinning shown by an arrow in Fig. 3(b).

### 3.3 Contact resistivity required for miniaturized graphene FETs

Finally, the $\rho_{C\square}$ value required for miniaturized graphene FETs is discussed. Consider that the condition on $\rho_{C\square}$ is such that the ratio of $R_C$ with $R_{ch}$ becomes less than 10 % because the FET performance should be mainly characterized by $R_{ch}$. It can be expressed by the following equation:

$$R_C/R_{ch} = \frac{\rho_{C\square}}{dW} \bigg/ R_{ch}^S \frac{L}{W} = \rho_{C\square}/R_{ch}^S dL = 0.1. \quad (3)$$

Figure 10 shows the $\rho_{C\square}$ required for $R_C/R_{ch} = 0.1$ as a function of $d$ for various $L$. In this calculation, a typical value for $R_{ch}^S = 250\ \Omega$ at $\mu = 5000\ \text{cm}^2\text{V}^{-1}\text{s}^{-1}$ and $n = 5\times10^{12}\ \text{cm}^{-2}$ was used. The dotted line that indicates the trace of $d_T$ for various $L$ was calculated by eq. (1). It separates the regions of "crowding" and "uniform injection". It is evident that the requirement of $\rho_{C\square}$ becomes more severe when the contact length $d$ becomes smaller than $d_T$. For a channel length of ~10 μm, the present status of $\rho_{C\square}$ for the Ni electrode satisfies the requirement. For the channel length of 100 nm, however, the required $\rho_{C\square}$ value is smaller than $10^{-9}\ \Omega\ \text{cm}^2$, which is four orders of magnitude lower than the present value. $d_T$ is on the order of 10 nm. This $\rho_{C\square}$ value is smaller than that required for the metal/Si contact (~$10^{-8}\ \Omega\ \text{cm}^2$) because the $R_{ch}^S$ of graphene is lower than that of Si. Recently, the high-frequency application of graphene has become a hot topic,[23] and the low $\rho_{C\square}$ required to achieve the higher transconductance is a critical issue.

To further decrease $\rho_{C\square}$ by four orders of magnitude, the factors that determine $\rho_{C\square}$ should be considered. As a first approximation, the work function difference ($\Delta\phi$) between graphene and metal is examined. The work functions of graphene, Ti, Cr, and Ni are 4.5, 4.3, 4.6, and 5.2 eV, respectively.[4,6,24] It is clear that Ni, which has the largest $\Delta\phi$, also has the lowest $\rho_{C\square}$, as shown in Fig. 5. In the case of a larger $\Delta\phi$, electrons are transferred from graphene to the metal, which will considerably increase the DOS in graphene under the metal contact and reduce $\rho_{C\square}$, as schematically shown in Fig. 2. Therefore, to obtain a low $\rho_{C\square}$, metals with larger $\Delta\phi$ are preferred.[12] However, $n$ doping of graphene on "clean" Ni(111) substrates has been reported,[25] which seems to be inconsistent with the above discussion. It is suggested that the orbital coupling between the π-band for graphene and the d-band for Ni results in this $n$ doping. This discrepancy is possibly caused by the nonrobustness of the fabrication processes at present. The organic resist residue is expected to be incorporated in the metal/graphene interface because it is difficult to remove it from the surface of graphene during the lithography process.[26]

### 4. Conclusions and Future Outlook

The contact resistance will be the performance killer in miniaturized graphene FETs; it needs to be lowered by several orders of magnitude from the present value of ~$5\times10^{-6}\ \Omega\ \text{cm}^2$. The systematic results suggest that metals with higher $\Delta\phi$ may be preferred to achieve low $\rho_{C\square}$ because of an increase in the DOS in graphene underneath the metal caused by the charge transfer. Elucidating the physical properties of graphene under the metal contact and the chemical interaction between graphene and metal is the key to realizing reliable contact with low $\rho_{C\square}$.

**Acknowledgements**

The Kish graphite used in this study was kindly provided by Dr. E. Toya of Covalent Materials Corporation. This work was partly supported by the Japan Society for the Promotion of Science (JSPS) through its "Funding Program for World-Leading Innovative R&D on Science and Technology (FIRST Program)" and by a Grant-in-Aid for Scientific Research from The Ministry of Education, Culture, Sports, Science and Technology, Japan.

**Figures**

**Fig. 1**     (Color online) Energy band diagrams for (a) metal/semiconductor, (b) metal/metal, and (c) metal/graphene contacts. $E_C$ and $E_V$ are the energies for the conduction and valence bands, respectively.

**Fig. 2**     (Color online) The relation between the DOS and the energy "just" at the metal/graphene interface (a) before the contact and (b) after the contact.

**Fig. 3**     (Color online) (a) A schematic of the graphene FET with the back gate. (b) A schematic of the band diagram showing the charge transfer region (hatched). The traces of the Dirac point for different $V_G$ are indicated by the broken line, the dash-dotted line, and the dash-double-dotted line. The respective resistivities are also shown as a function of position (c). (d) A schematic of the conductivity (σ) curves for the cases (left) without the charge transfer and (right) with the charge transfer. An asymmetric σ curve is observed because of the additional resistance produced by the p-n junction.

**Fig. 4**     (Color online) (a) An optical micrograph of the four-layer graphene device with six sets of four-probe configurations (#1~#6). The contact metal is Ni. (b) A schematic of the device. (c) Two types of contact resistivity, $R_CA$ and $R_CW$, extracted by a four-probe measurement from the devices in (a). The unit for $\rho_C=R_CA$ is Ω μm$^2$, while it is Ω μm for $\rho_C=R_CW$.

**Fig. 5**     (Color online) Contact resistivities ($\rho_C=R_CW$) for the contact metals Cr/Au, Ti/Au and Ni as a function of $\mu_{4P}/\mu_{2P}$. The colors indicate the layer number.

**Fig. 6** (Color online) A schematic of the transmission line model for the metal/graphene contact.

**Fig. 7** (Color online) (a) An optical micrograph of the cross-bridge Kelvin structure for monolayer graphene with a rectangular shape. (b) A schematic of the electric potential along the dotted line in (a). (c) $\rho_{C\square}$ and $d_T$ as a function of the gate voltage.

**Fig. 8** (Color online) $\rho_C$ as a function of $V_G$ extracted by the four-probe measurement for the Ni contact.

**Fig. 9** (Color online) Schematics of 3 types of $\rho_C$–$V_G$ relations. (a) The relative positions of the Dirac points for $R_{Total}$ and $R_{ch}$ are different (a) and almost equal (b).

**Fig. 10** (Color online) The $\rho_{C\square}$ required for $R_C/R_{ch} = 0.1$ as a function of $d$ for various $L$. In this calculation, a typical value for $R_{ch}^S = 250\ \Omega$ at $\mu = 5000\ \mathrm{cm^2V^{-1}s^{-1}}$ and $n = 5\times10^{12}\ \mathrm{cm^{-2}}$ was used. $\rho_{C\square}$ is constant for $d > d_T$ (crowding), while it decreases for $d < d_T$ (uniform injection).

**Fig. 1**

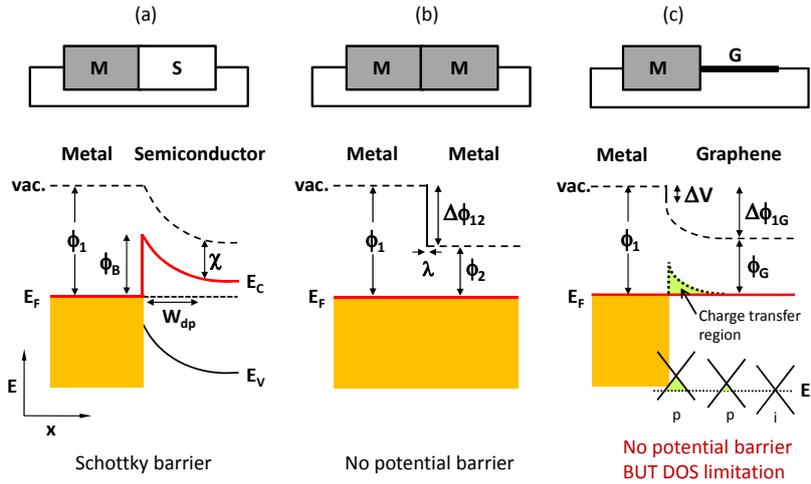

**Fig. 2**

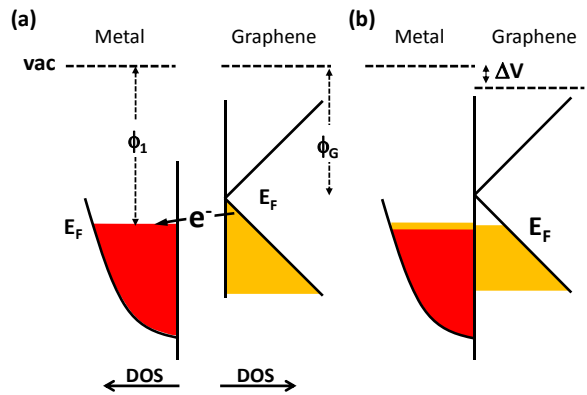

**Fig. 3**

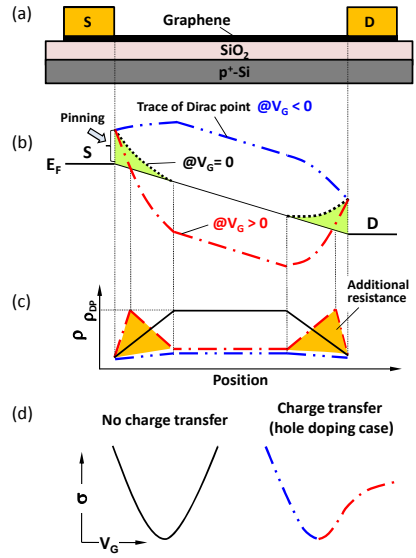

**Fig. 4**

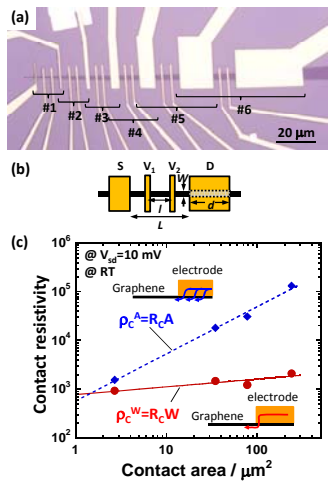

**Fig. 5**

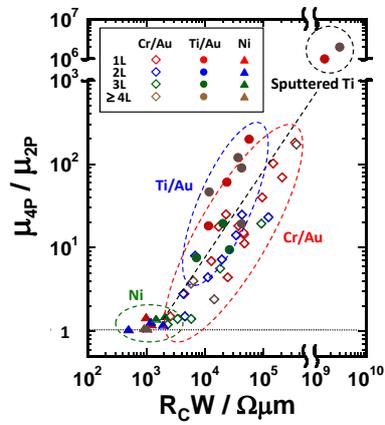

**Fig. 6**

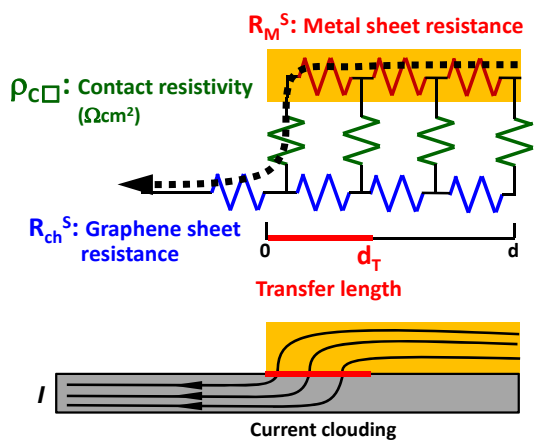

**Fig. 7**

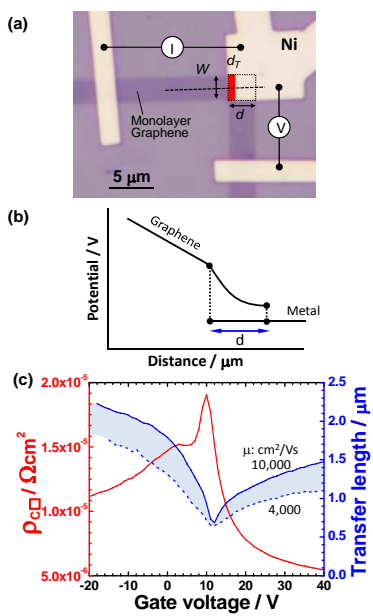

K. Nagashio et al.

**Fig. 8**

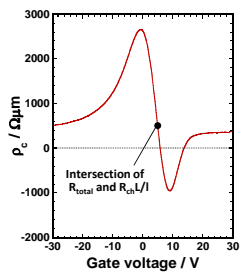

**Fig. 9**

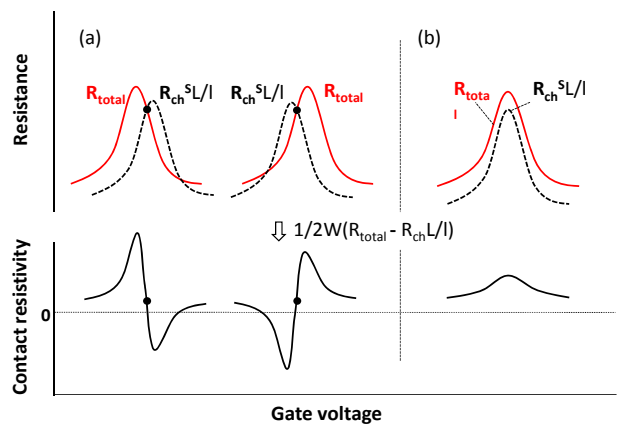

**Fig. 10**

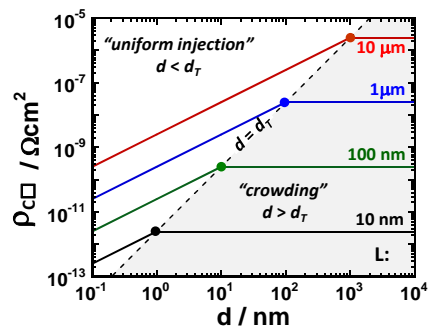